\documentclass{kapproc} 

\usepackage{procps}

\usepackage[dvips]{graphicx}

\upperandlowercase

\normallatexbib 
\begin{document}

\articletitle{Prediction Possibility in the\\ 
 Fractal Overlap Model of Earthquakes}


\author{Srutarshi Pradhan}
\affil{Saha Institute of Nuclear Physics\\
1/AF Bidhan Nagar, Kolkata 700064, India}
\email{spradhan@cmp.saha.ernet.in}

\author{Pinaki Choudhuri}
\affil{Department of Physics\\
Indian Institute of Science, Bangalore 560012, India}
\email{pinakic@physics.iisc.ernet.in}
\author{Bikas K. Chakrabarti}
\affil{Saha Institute of Nuclear Physics\\
1/AF Bidhan Nagar, Kolkata 700064, India}
\email{bikas@cmp.saha.ernet.in}

\begin{abstract}

\noindent The two-fractal overlap model of earthquake shows that 
the contact area distribution of two fractal surfaces follows power law decay
in many cases and this agrees with the Guttenberg-Richter power law. Here, we
attempt to predict the large events (earthquakes)  in this model through 
the overlap time-series analysis. Taking only the Cantor sets, the overlap 
sizes (contact areas) are noted  when one Cantor set  moves
over the other with uniform velocity. This  gives a time series  containing 
different overlap sizes. Our numerical study here shows that 
the cumulative overlap size grows almost linearly with time and when the overlap
sizes are added up to a pre-assigned large event (earthquake) and then reset 
to `zero' level, the corresponding  cumulative overlap sizes grows upto 
some  discrete 
(quantised) levels.  This observation should help to predict the 
possibility of `large events' in this (overlap) time series. 
\end{abstract}


\section{Introduction}
\noindent The two-fractal overlap model of earthquake  \cite{BS99} is a 
very recent modeling attempt. 
Such  models are all  based on the observed `plate tectonics'
and the fractal nature of the interface between tectonic plates 
and earth's solid crust.  
The statistics of overlaps between two fractals is 
not studied much yet, though their knowledge is often required in various 
physical contexts. For example, it has been established recently that since 
the fractured surfaces have got well-characterized self-affine properties
\cite{BB82,BB84,Hansen03}, the distribution of the elastic energies
released during the slips between two fractal surfaces (earthquake
events) may follow the overlap distribution of two self-similar fractal
surfaces \cite{BS99,SBP03} (see also \cite {VD96}). Chakrabarti and 
Stinchcombe \cite{BS99}
had shown analytically that for regular fractal overlap (Cantor sets
and carpets) the contact area distribution follows a simple power
law decay.

The two fractal overlap magnitude changes in time as one fractal moves
over the other. The overlap (magnitude)
time series can therefore be studied as a model time series of earthquake
avalanche dynamics \cite{BK67,CL89}.

Here, we study the time (\( t \)) variation of contact area (overlap)
 \( m(t) \) between two well-characterized fractals having the same
fractal dimension as one fractal moves over the other with constant
velocity. We have chosen only very simple fractals: regular or non-random
Cantor sets and random Cantor sets. We analyse the time 
series data of Cantor set overlaps to find the prediction possibility 
of large events (occurrence of large overlaps). we show that the time 
series \( m(t) \) obtained by moving one fractal uniformly over the 
other (with periodic boundary condition) has some features which can be 
utilized to predict the
`large events'.  This is shown utilizing
the discrete or a `quantum' nature of the integrated (cumulative) 
overlap over time.

\section{The fractal overlap model}
\noindent We consider here the overlap sets of two identical fractals,
as one slides over the other. We study the overlap distribution \( P(m) \)
for the overlap set \( m \) defined at various (sliding) positions
of the two fractals. First, we consider two regular cantor sets 
(at finite generation \( n \)).

\resizebox*{5cm}{3cm}{\includegraphics{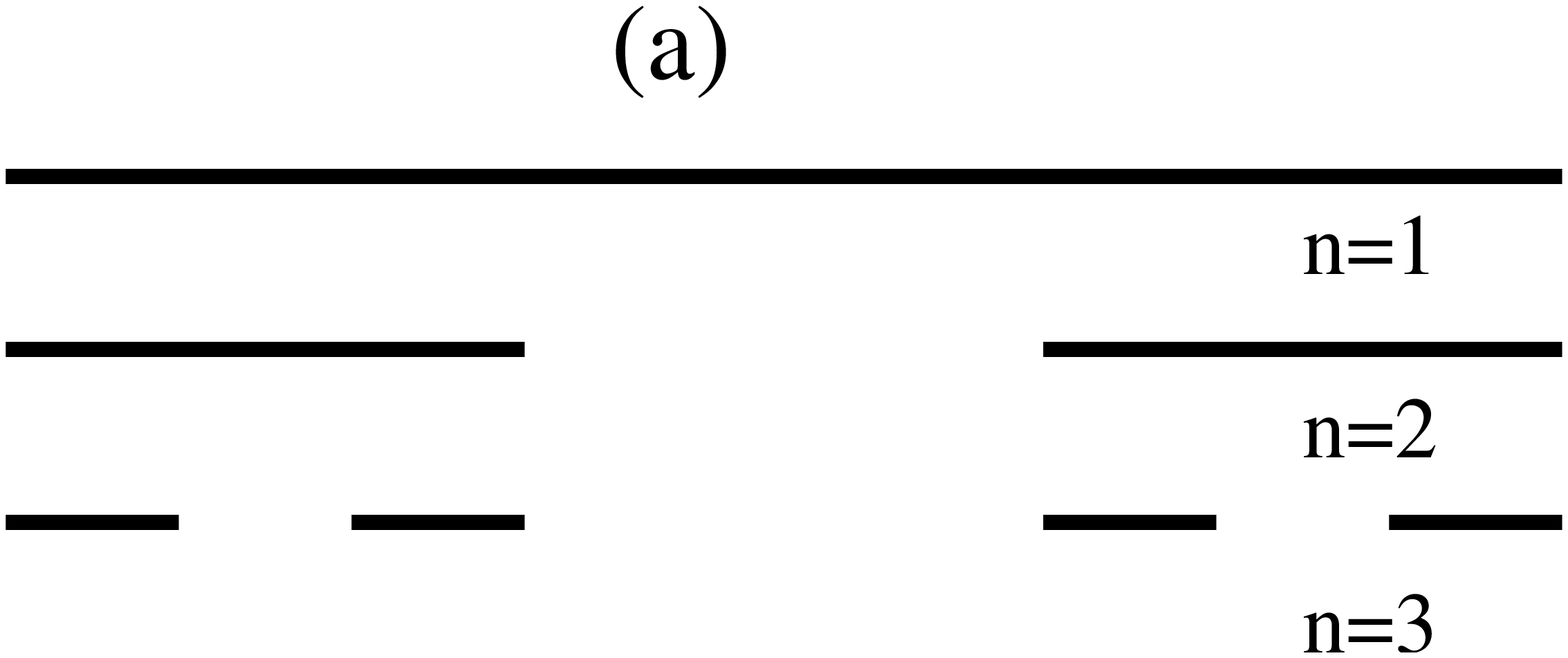}} \hskip.3in\resizebox*{5cm}{3cm}{\includegraphics{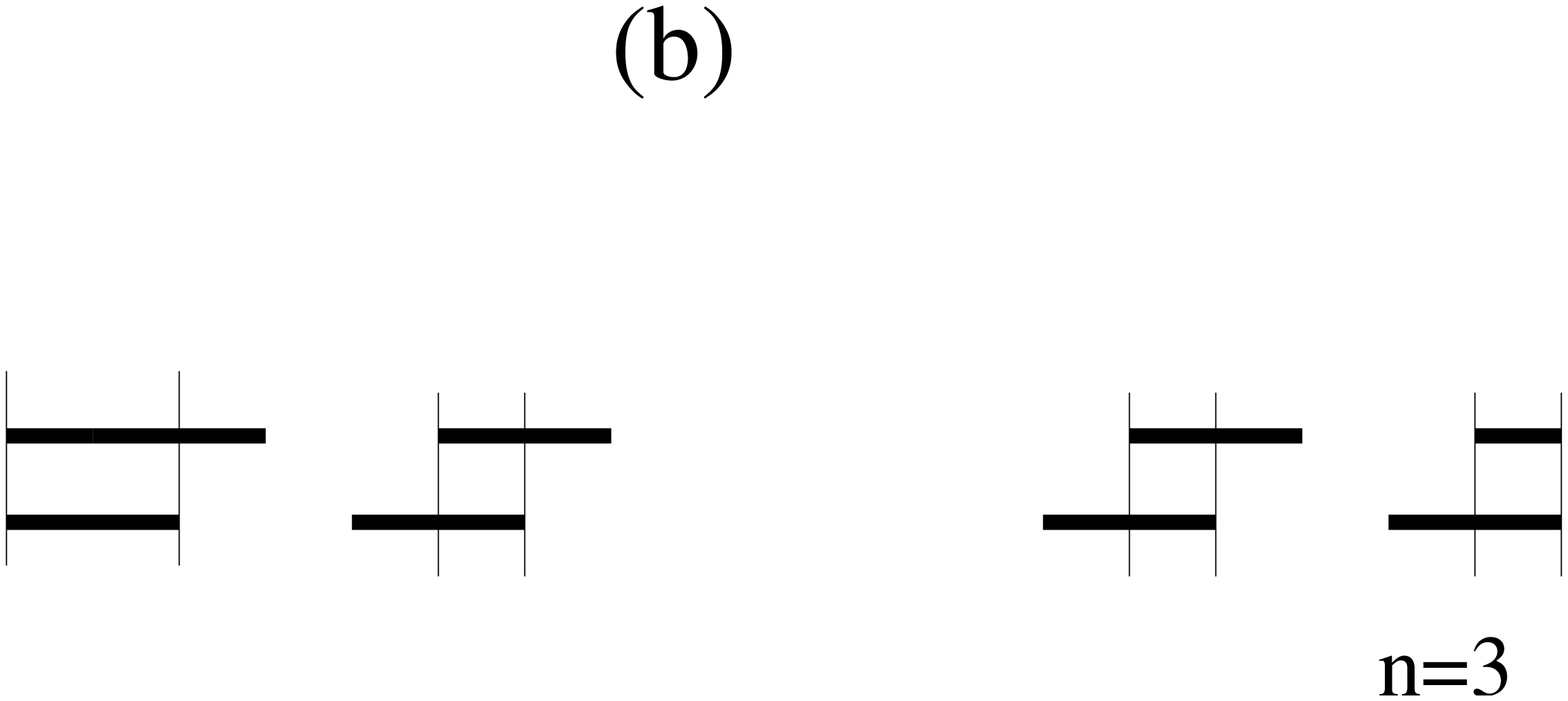}}

\vskip.1in


\vskip.1in

\noindent \textbf{\small \it Figure 1.} {\small (a) A regular Cantor set 
of dimension
\( \ln 2/\ln 3 \); only three finite generations are shown. (b) The
overlap of two identical (regular) Cantor sets, at \( n=3 \), when
one slides over other; the overlap sets are indicated within the vertical
lines, where periodic boundary condition has been used.} 

{\small \vskip.2in}{\small \par}



\noindent Next, we consider the same for two identical random cantor
sets  (at finite generations \( n \)).

\vskip.1in
\resizebox*{5cm}{4cm}{\includegraphics{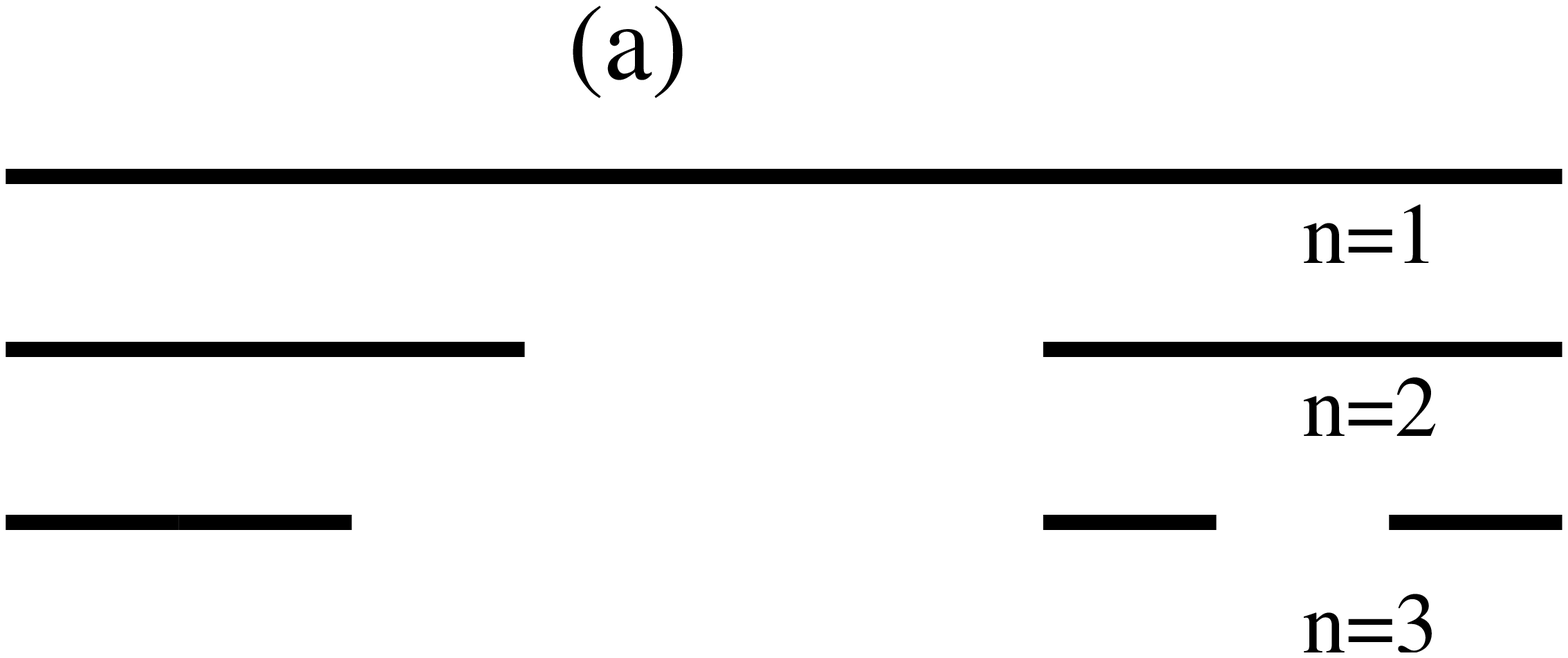}} \hskip.3in\resizebox*{5cm}{4cm}{\includegraphics{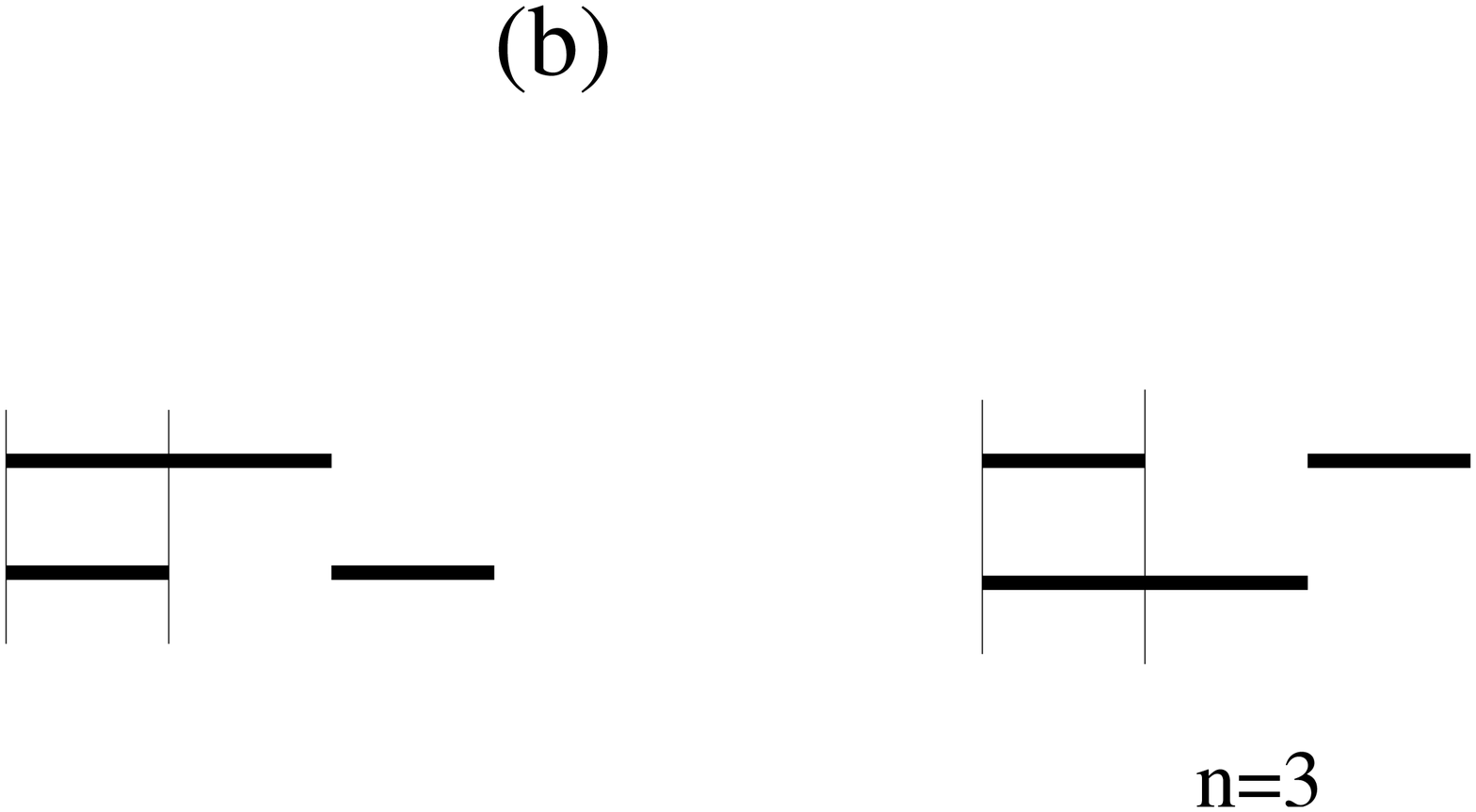}}

\vskip.1in


\vskip.1in

\noindent\textbf{\small \it Figure 2.} {\small (a) A random Cantor set of 
dimension
\( \ln 2/\ln 3 \); only three finite generations are shown. (b) Overlap
of two random Cantor sets (at \( n=3 \); having the same fractal
dimension) in two different realisations. The overlap sets are indicated
within the vertical bars.} 

\vskip .1in
We studied earlier \cite {SBP03} the overlap statistics for regular and random 
Cantor sets, gaskets and percolating clusters \cite {Stauffer92}. 
We found a universal scaling behavior of the overlap
or contact area (\( m \)) distributions \( P(m) \) for all types of fractal set 
overlaps mentioned:  $P^{\prime }(m^{\prime })=L^{\alpha }P(m,L);
m^{\prime }=mL^{-\alpha },$
where \( L \) denotes the finite size of the fractal and the exponent
\( \alpha =2(d_{f}-d) \); \( d_{f} \) being the mass dimension of
the fractal and \( d \) is the embedding dimension. Also the overlap
distribution \( P(m) \), and hence the scaled distribution \( P^{\prime }(m^{\prime }) \),
are seen to decay with \( m \) or \( m^{\prime } \) following a power
law (with exponent value equal to the embedding dimension of the fractals)
for both regular and random Cantor sets and gaskets:
$P(m)\sim m^{-\beta };\beta =d.$
However, for the percolating clusters \cite{Stauffer92}, the overlap size 
distribution takes a Gaussian form. 

\section{Time series  analysis for two Cantor set overlaps}
\noindent We consider now the time series obtained by counting the
overlaps \( m(t) \) as a function of time as one Cantor set moves
over the other (periodic boundary condition is assumed) with uniform velocity.

\subsection{ Overlap time series data}

\noindent The time series are shown in Fig. 3., for finite generations
of Cantor sets of dimensions \( \ln 2/\ln 3 \) and \( \ln 4/\ln 5 \)
respectively.

\resizebox*{5.5cm}{5cm}{\includegraphics{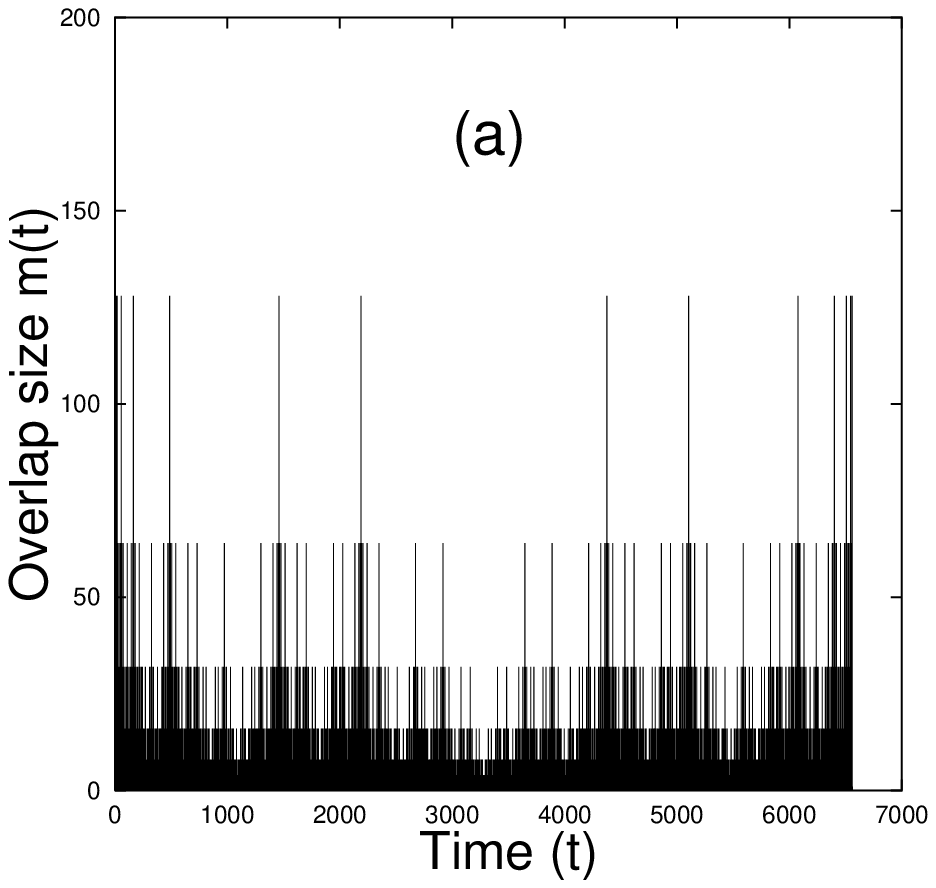}} \hskip.2in\resizebox*{5.5cm}{5cm}{\includegraphics{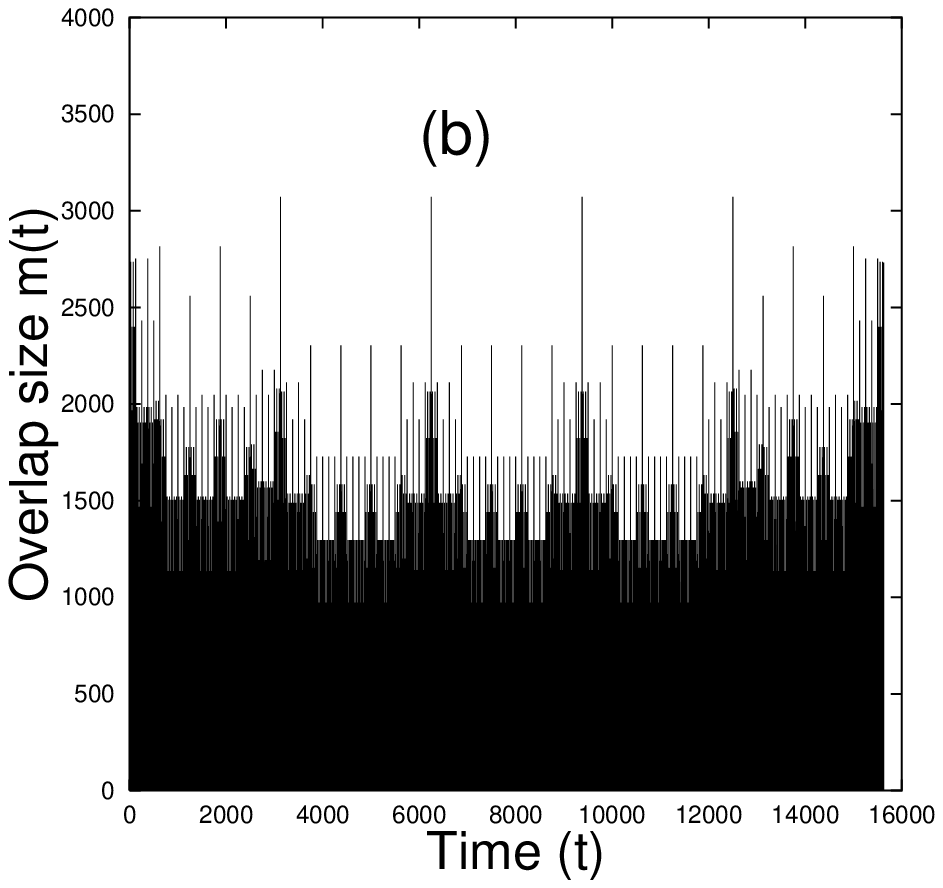}}

\vskip.1in
\noindent\textbf{\small \it Figure 3.} {\small The time (\( t \)) series data of
overlap size (\( m \)) for regular Cantor sets: (a) of dimension
\( \ln 2/\ln 3 \), at \( 8 \)th generation: (b) of dimension \( \ln 4/\ln 5 \),
at \( 6 \)th generation. The obvious periodic repeat of the time series 
comes from the assumed periodic boundary condition of one of the 
sets (over which the other one slides).}{\small \par}

\subsection{Cumulative overlap sizes }

\noindent Now we calculate the cumulative overlap size
$Q(t)=\int _{o}^{t}mdt$
\noindent and plot that against time in Fig. 4. Note, that `on average'
\( Q(t) \) is seen to grow linearly with time \( t \) for regular
as well as random Cantor sets. This gives a clue that instead of looking
at the individual overlaps \( m(t) \) series one may look for the
cumulative quantity. In fact, for the regular Cantor set of dimension
\( \ln 2/\ln 3 \), the overlap \( m \) is always \( 2^{k} \), where
\( k \) is an integer. However the cumulative \( Q(t)=\sum _{i=0}^{t}2^{k_{i}} \)
can not be easily expressed as any simple function of \( t \). Still,
we observe \( Q(t)\simeq ct \), where \( c \) is 
dependent on the fractal. This result is even more prominant in the case 
of Cantor sets with \( d_{f}= \) \( \ln 4/\ln 5 \).

\vskip.2in
\resizebox*{5cm}{4.5cm}{\includegraphics{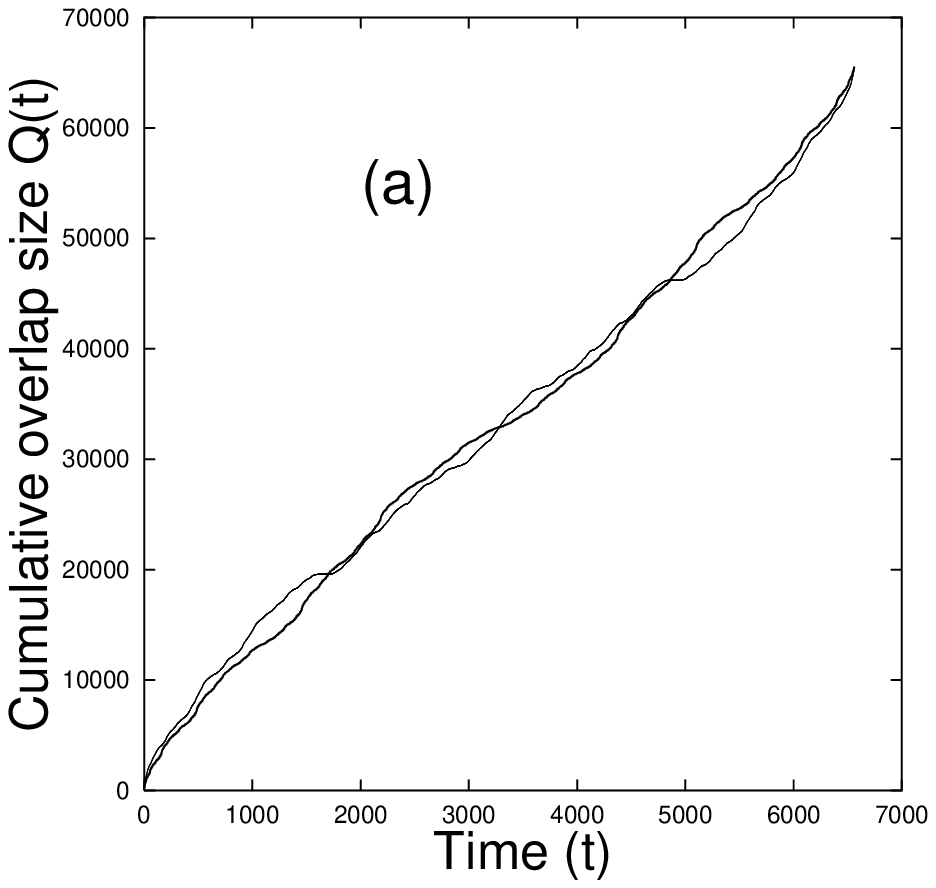}} \hskip.2in\resizebox*{5cm}{4.5cm}{\includegraphics{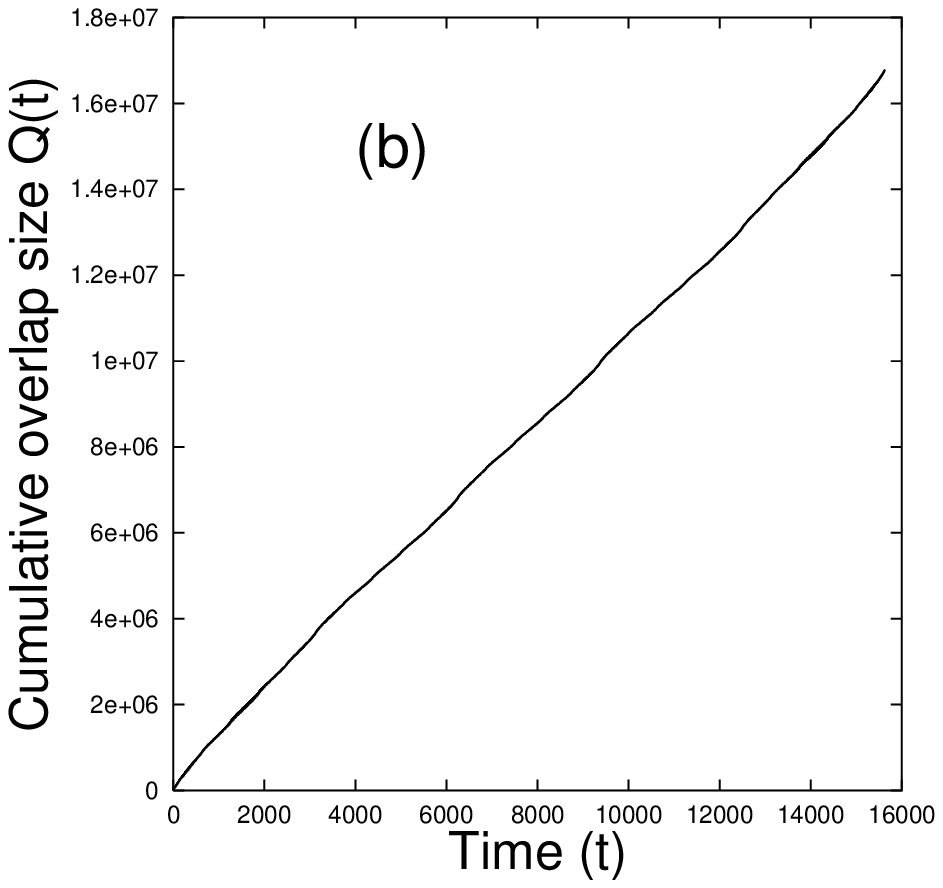}}

\vskip.1in
\noindent\textbf{\small \it Figure 4.} {\small The cumulative overlap \( Q \) versus
time; for pure cantor sets: (a) of dimension \( \ln 2/\ln 3 \) (at
\( 8 \)th generation) and (b) of dimension \( \ln 4/\ln 5 \) (at
\( 6 \)th generation). The dotted line corresponds to those for two identical
but random Cantor sets. In (b) the two lines fall on each other.}{\small \par}

\subsection{ Cumulative overlap quantization}

\noindent We first identify the `large events' occurring at time \( t_{i} \)
in the \( m(t) \) series, where \( m(t_{i})\geq M \), a pre-assigned
number. Then, we look for the cumulative overlap size 
\( Q_{i}(t)=\int _{t_{i-1}}^{t}mdt \), \( t\) \( \leq t_{i}\),
where the successive large events occur at times \( t_{i-1} \) and
\( t_{i} \). The behavior of \( Q_{i} \) with time is shown in Fig.
5 for regular cantor sets with \( d_{f}=  \) \( \ln 2/\ln 3 \) at
generation \( n=8 \). Similar results are also given for Cantor sets
with \( d_{f}=  \) \( \ln 4/\ln 5 \) at generation \( n=6 \) in Fig.
6. \( Q_{i} (t)\) is seen to grow almost lonearly in time upto \(Q_{i}(t_{i}\)) 
after which it drops down to zero. It appears that there are discrete 
(quantum) values of \( Q_{i} (t_{i}\)).

\resizebox*{5.5cm}{5.5cm}{\includegraphics{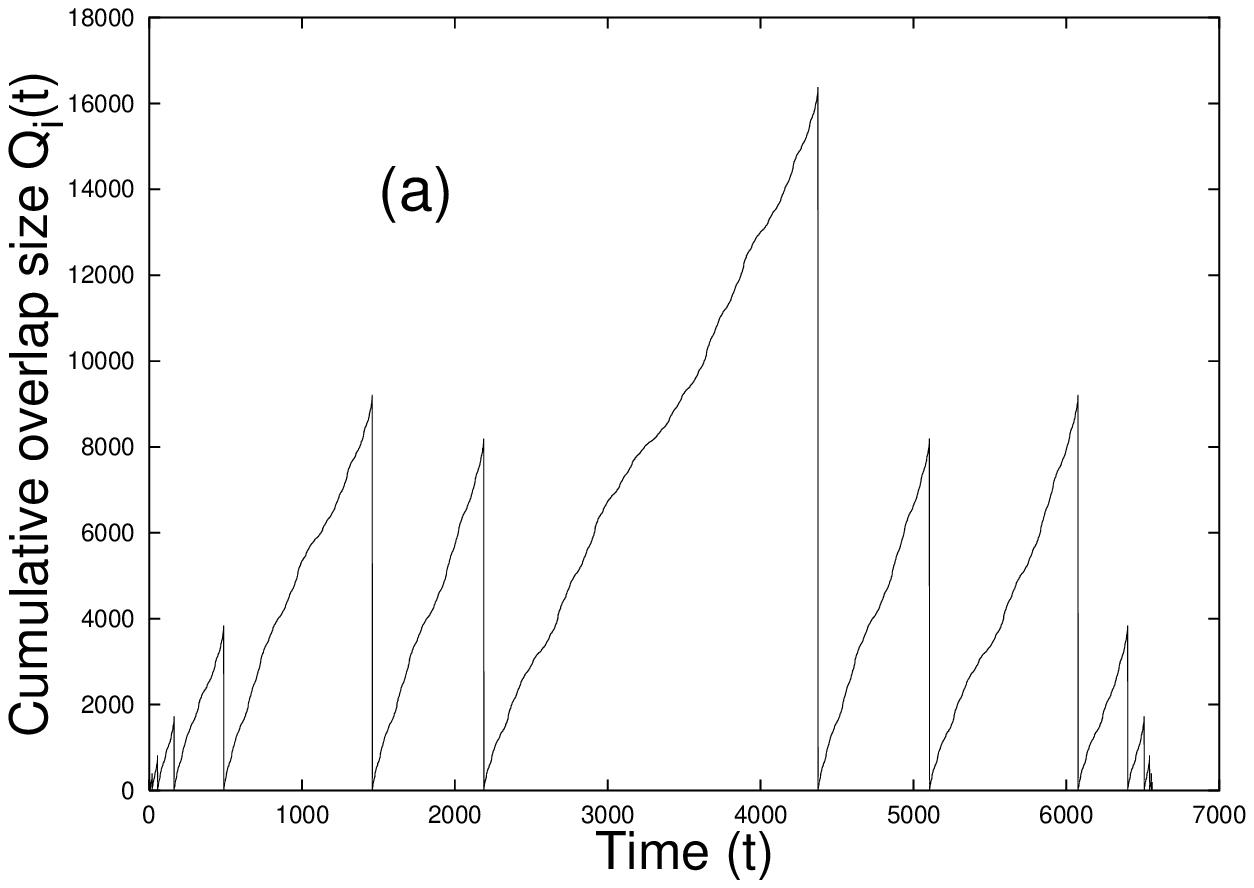}} \hskip.3in\resizebox*{5.5cm}{5.5cm}{\includegraphics{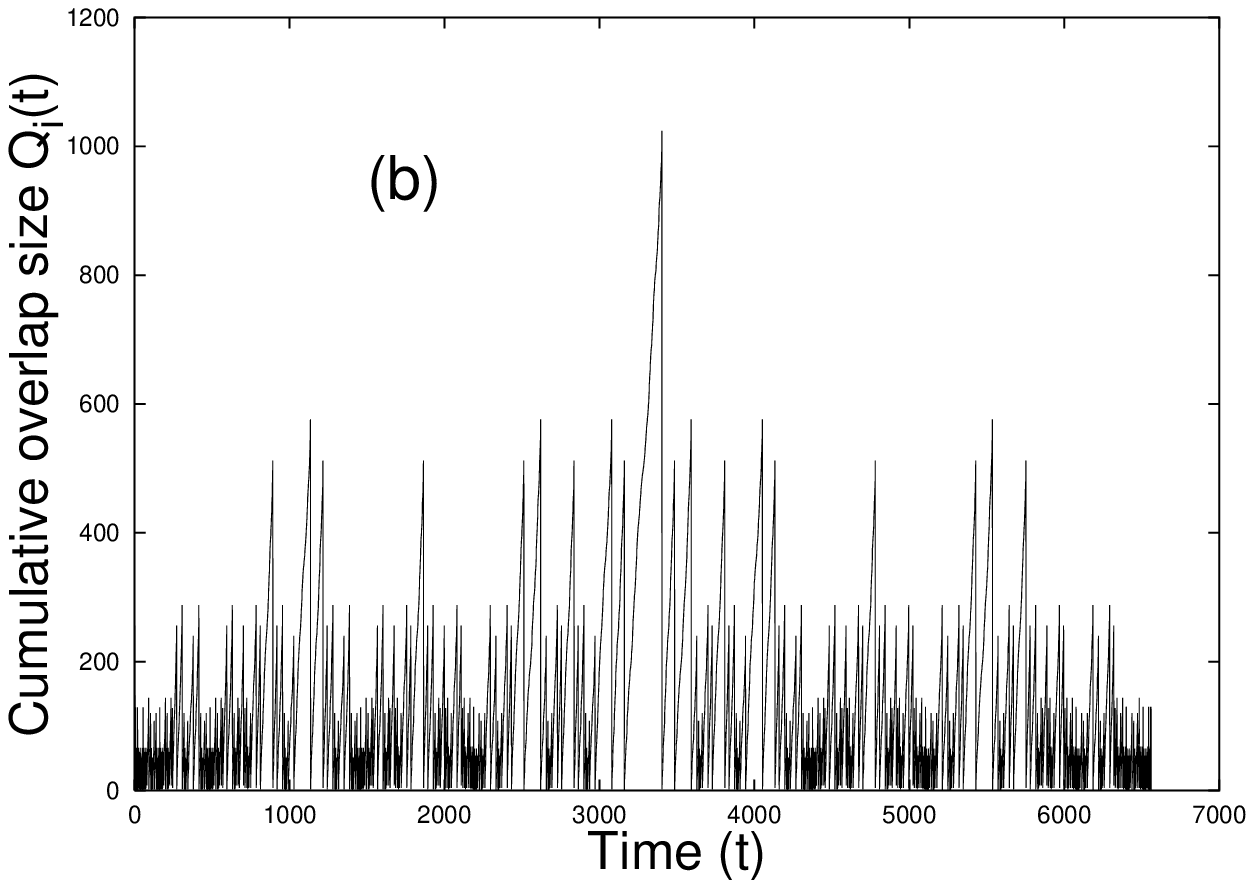}}

\vskip.1in

\noindent\textbf{\small \it Figure 5.} {\small The cumulative overlap 
size variation
with time (for regular Cantor sets of dimension \( \ln 2/\ln 3 \),
at \( 8 \)th generation), where the cumulative overlap has been reset
to \( 0 \) value after every big event (of overlap size \( \geq M \)
where \( M=128\) (a) and \( 32 \) (b) respectively).}

\resizebox*{5.5cm}{5.5cm}{\includegraphics{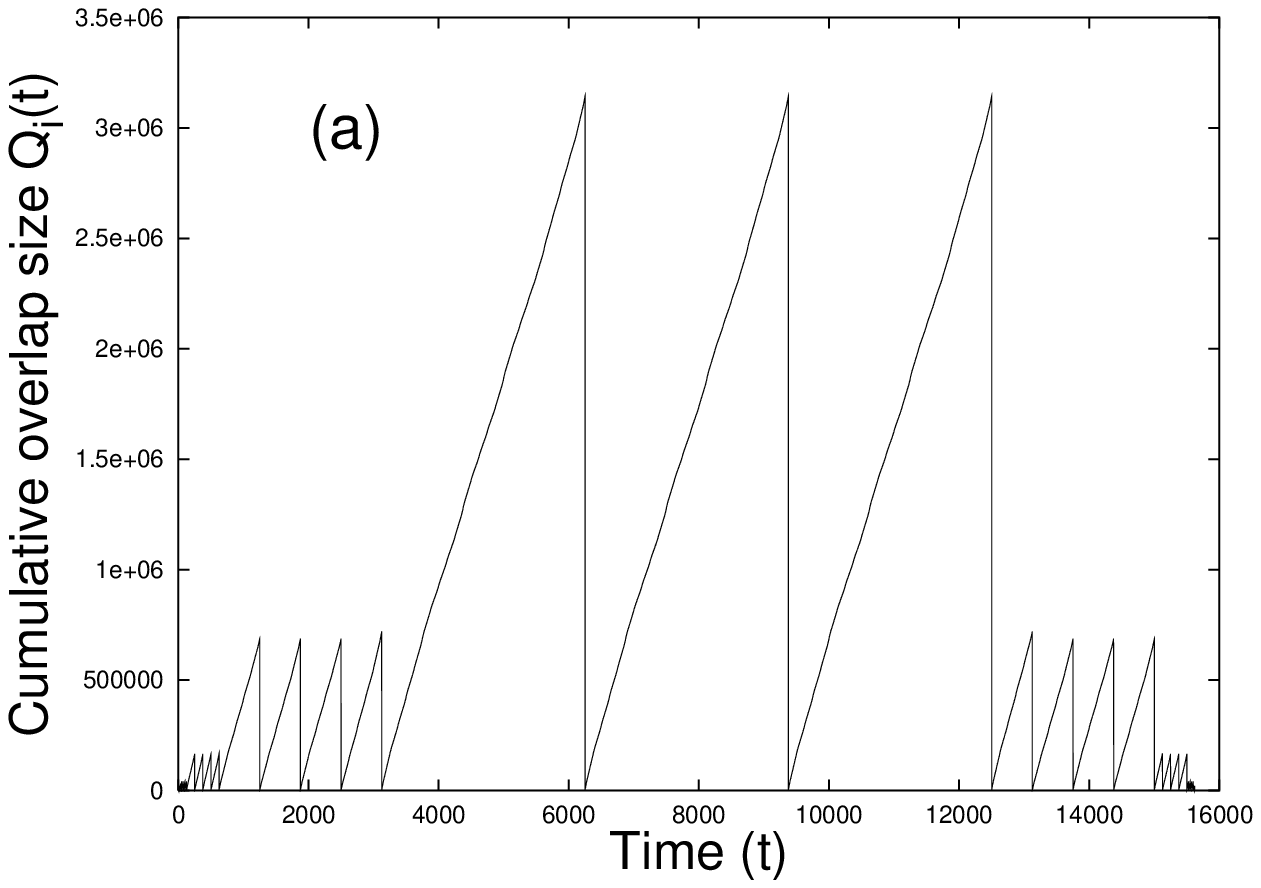}} \hskip.2in\resizebox*{5.5cm}{5.5cm}{\includegraphics{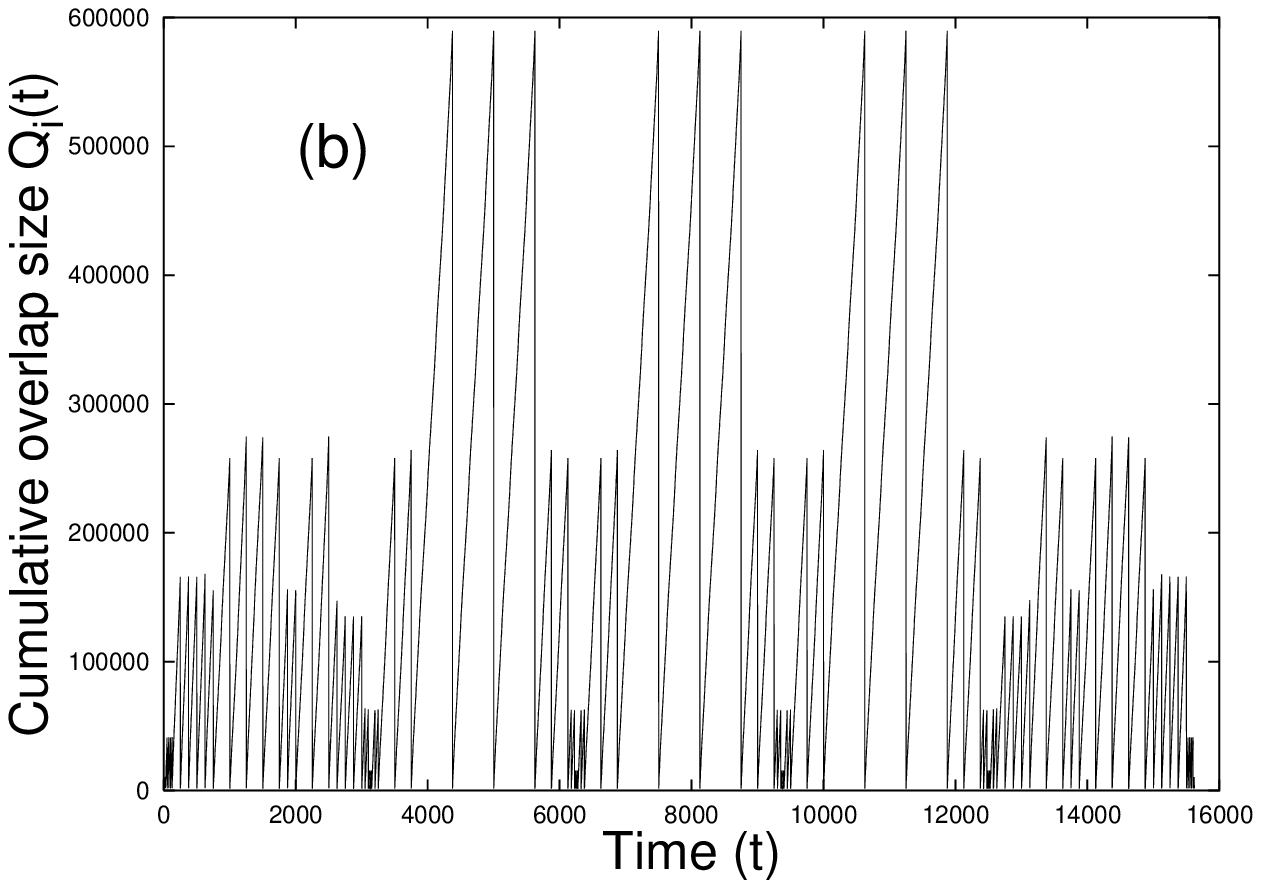}} 

\noindent\textbf{\small \it Figure 6.} {\small The cumulative overlap 
size variation
with time (for regular Cantor sets of dimension \( \ln 4/\ln 5 \),
at \( 6 \)th generation), where the cumulative overlap has been reset
to \( 0 \) value after every big event (of overlap size \( \geq M \)
where \( M=2400\) (a) and \(2048 \) (b) respectively). }{\small \par}

\section{Summary and discussion}

\noindent If one Cantor set moves uniformly over another, the overlap
between the two fractals change quasi-randomly with time (see eg.,
Fig. 3). The overlap size distribution was argued \cite{BS99} and
shown \cite{SBP03} to follow power law decay. Here we show numerically that if
one fixes a magnitude \( M \) of the overlap sizes \( m \), so that
overlaps with \( m\geq M \) are called `events' (or earthquake),
then the cumulative overlap \( Q_{i}(t)=\int ^{t}_{t_{i-1}}mdt \), 
\(t\) \( \leq t_{i} \), 
(where two successive events of \( m\geq M \) occur at times \( t_{i-1} \)
and \( t_{i} \)) grows linearly with time up to some discrete quanta
\( Q_{i}(t_{i})\cong lQ_{0} \), where \( Q_{0} \) is the minimal overlap
quantum, dependent on \( M \) and \( n \) (the generation number). 
Here \( l \) is an integer (see Figs. 5, 6). Although our results here are 
for regular fractals
 of finite generation \( n \), the observed discretisation of the 
overlap cumulant \( Q_{i} \)  with the time limit set by \( n \), is a robust 
feature and can be seen for larger time series for larger generation
 number \( n \).  Similar studies for random Cantor set overlap are in progress.
This model study therefore indicates that one can 
note the growth of the cumulant sesmic response \( Q_{i} (t) \),
rather than the sesmic event strength \( m(t) \), and anticipate some big 
events as the response reaches the discrete levels \( lQ_{0} \), specific to 
the series of events.  
 

\begin{acknowledgments}
\noindent We would like to thank G. Ananthakrishna and P. Bhattacharyya 
for some useful discussions. 
\end{acknowledgments}

\end{document}